\documentclass[12pt,fleqn]{article}

\newcommand{\be}{\begin{eqnarray}}
\newcommand{\ee}{\end{eqnarray}}
\newcommand{\trg}{\mbox{trg}\,}
\newcommand{\D}{\Delta}
\newcommand{\Si}{\Sigma}
\usepackage{graphicx}

\begin{document}
 \title{Thouless energy in QCD and effects of diffusion modes
 on level correlations of Dirac operator}
 \author{K. Takahashi$^a$ and S. Iida$^b$}
 \date{{\small{\it $^a$Institute of Physics, University of Tsukuba,
 Tsukuba 305-8571, Japan}\\
{\it $^b$Faculty of Science and Technology,  Ryukoku University,
 Otsu 520-2194, Japan}}}
 \maketitle

 \begin{abstract}
 The correlations of the QCD Dirac eigenvalues are studied with use of 
 an extended chiral random matrix model. 
 The inclusion of spatial dependence which the original model lacks 
 enables us to investigate the effects of diffusion modes.
 We get analytical expressions of level correlation functions with 
 non-universal behavior caused by diffusion modes
 which is characterized by Thouless energy.
 Pion mode is shown to be responsible for these diffusion effects 
 when QCD vacuum is considered a disordered medium.
 \end{abstract}
 \section{Introduction}
 It is now widely recognized that a chiral random matrix theory (ChRMT)
 \cite{sv} is a useful theoretical tool to describe
 non-perturbative aspects of QCD.
 Like other complicated quantum systems\cite{rmt}, statistical properties of 
 QCD Dirac operator show high universality described by this model.
 A possible relation between chiral phase transition and 
 Anderson localization transition (a metal-insulator transition induced by
 a random impurity potential)
 \footnote{There are some arguments that Anderson transition can only happen 
 in quenched theories (see e.g. Ref.\cite{rep})}, 
 which was originally suggested 
 in the context of an instanton liquid model\cite{ins1,ins2}, 
 is currently under intensive study.

 In order to discuss such relations, however, the original model is 
 not appropriate because the model lacks space-time coordinate dependence 
 and can only describe a `metallic' region.
 In this region, energy eigenfunctions extend over the whole system.
 The correlations of energy levels are very large and can be correctly
 described by RMT.
 But, if impurity potential becomes strong,
 the wavefunction tends to localize.
 The correlations become small and the diffusion effects 
 become important.

 Recently, some authors have discussed the limitation of 
 the original model\cite{chdis,Th}. 
 Comparing with numerical simulations,
 they found that the deviations from the results of
 the original model could be seen beyond a certain energy scale\cite{Th,lat}. 
 Further there were some analytical arguments supporting the above 
 results\cite{pqch,univ}.

 This energy scale is termed Thouless energy after  
 a quantity with a similar role in Anderson problem. 
 In  Anderson problem, Thouless energy $E_c$ is defined as
 $D/L^{2}$ where $D$ is a diffusion constant and $L$ is a system size. 
 If energy scale we consider is much smaller than $E_c$,  
 the spatial structure of a system can be ignored and
 we can use a random matrix model with a proper symmetry as
 a good approximation to describe statistical properties of a system.
 But once energy scale becomes comparable to or larger than $E_c$, 
 we can no longer ignore the spatial dependence of the system.

 The intuitive explanation of above results is as follows;
 Since a typical distance of a diffusing particle with
 a diffusion constant $D$ during a time period $t$ is $\sqrt{Dt}$,
 a time scale in which a particle is diffusing 
 throughout a sample with linear dimension  $L$ is $\tau_c = L^{2}/D$. 
 The corresponding energy scale, $1/\tau_{c}$, is Thouless energy.
 Hence, when we consider times much greater than $\tau_c$
 (or energies much smaller than $E_c$),
 a particle has enough time to wander everywhere in the sample 
 and would finally forget from where it starts.
 Then every point in the sample becomes equivalent and 
 the spatial structure of the original system can be ignored.

 Besides the above intuitive reasoning, Thouless energy also appears 
 in expressions of various statistical quantities in Anderson problem.
 In the context of impurity perturbation technique, 
 Thouless energy is defined as a first excitation energy of fundamental 
 diffusion modes, diffuson.  
 If all excitation energies of the modes with finite wave numbers are much 
 greater than the energy scale we consider,
 we can ignore contributions from these modes.
 There remains the lowest mode with 0 wave number 
 which is constant throughout the sample. 
 Keeping only the lowest mode amounts to neglecting the spatial dependence. 

 Turning to the statistical properties of QCD Dirac operator, 
 the authors of Ref.\cite{chdis,Th} followed the above intuitive argument 
 and suggested that Thouless energy in QCD  is 
 the right hand side of Eq.(\ref{Ec}).
 Hence if relevant energy scale exceeds this quantity, we should expect 
 deviations from ChRMT due to the effects of diffusion modes. 
 At present, however, we do not know concrete expressions of these 
 effects, which is necessary to make the arguments quantitative.
 
 In this paper, we aim to go beyond the qualitative argument 
 and to calculate these effects of diffusion analytically.
 Guided by a model used to describe disordered electrons in
 metals\cite{iwz}, we extend the original chiral random matrix model;
 a single random matrix is replaced by 
 a set of mutually coupled random matrices 
 each of which is localized in a finite space-time region.
 This model is defined in section 2.
 Partition function is calculated and comparison with
 the nonlinear sigma model is made.

 To calculate the level density in RMT, one usually use
 the orthogonal polynomial method\cite{mehta} or
 the supersymmetry method\cite{ef,vwz}.
 In the present paper, we use the supersymmetry method because
 this method is easy to handle the extended model. 
 This is formulated in section 3.

 The averaged level density (section 4) and 
 the two-point level correlation function (section 5) are 
 calculated perturbatively.
%
 In comparison with the non-chiral RMT, we confirm that
 the Thouless energy in QCD is related to the pion decay constant 
 as originally suggested.

 In section 6, we also calculate the two-point correlation functions
 which correspond to meson propagators. 
 We find that a pion propagator derived in the present model 
 corresponds to a diffuson propagator in Anderson model. 
 Namely, pion mode plays the role of diffusion mode
 in disordered QCD vacuum.
 \section{Chiral random matrix model}
 \subsection{Original chiral random matrix model}
 Original chiral random matrix theory is defined as follows \cite{sv}.
 Generally, Euclidean Dirac operator matrix in gauge field can be written as
\be
 H&=&\left( \begin{array}{cc} 0 & W \\  W^{\dag} & 0 \end{array}\right)
 \nonumber\\
 &=&\left( \begin{array}{cc} 0 & \omega_1-i\omega_2 \\
 \omega_1+i\omega_2  & 0 \end{array}\right) \label{dirac}
\ee
 where $W$ is an $N\times N$ complex matrix and $\omega_{i}$ is a real matrix.
 That is, we consider the case of unitary ensemble
 which has the symmetry for fundamental fermions with three colors.
 N goes to infinity in the thermodynamic limit.
 Matrix $W$ depends on the gauge field and treated as random matrix.
 Gauge field path integral is replaced by Gaussian ensemble average\cite{sv} as
\be
 Z_{RMT}=\int dW\prod_{f=1}^{N_f}\mbox{det}\left(
   \begin{array}{cc}
       m_f     &  iW   \\
    iW^{\dag}  &  m_f
   \end{array}\right)
   \exp\left(-\frac{N}{2}\Si^2\mbox{Tr}WW^{\dag}\right)
\ee
 where $m_f$ are quark masses and $\Si$ denotes chiral condensate.
 This function is expressed by the integral of auxiliary field $U$ 
 which is an $N_f \times N_f$ unitary matrix\cite{sv} as
\be
 Z_{RMT}=\int dU\exp\left(N\Si\,\mbox{Re}\,\mbox{Tr}\,(m_qU)\right).
\ee
 This function is the same as the partition function of nonlinear sigma model
 without space-time dependence,
\be
 Z_{NLS}&=&\int [dU]\exp\left(-\int d^4x{\cal L}\right)\\
 &\to& \int dU\exp\left(V\Si\,\mbox{Re}\,\mbox{Tr}\,(m_qU)\right)\\
  & & {\cal L}=\frac{f_{\pi}^{2}}{4}
 \mbox{Tr}\left(\partial_{\mu}U^{\dag}(x)\partial_{\mu}U(x)\right)
 -\Si\mbox{Re}\left(\mbox{Tr}\,m_qU(x)\right)+\cdots
\ee
 where $f_{\pi}$ is a pion decay constant.
 In the R.H.S. of the arrow, we ignore the space time dependence of $U(x)$.
 That is, ChRMT is a zero dimensional theory.
 In ChRMT, we focus our attention on quasi-zero mode levels of Dirac operator.
 As long as we consider these quasi-zero modes, we can use the
 partition function of ChRMT without spatial dependence.
 It is known that the microscopic spectral density whicn is 
 the level density of  quasi-zero modes does not depend on 
 the assumption of Gaussian ensemble average \cite{admn}.

 In nonlinear sigma model, the condition for the zero-momentum sector 
 to dominate is $1/m_{\pi}\gg L$ \cite{ls} where $m_{\pi}$ is 
 the pion mass and $L$ is a  system size.
 This condition means Pion compton wavelength is greater than the system size.
 When $1/m_{\pi}\sim L$, contributions from nonzero modes are important
 and we cannot neglect the kinetic term.
 Using Gell-mann--Oakes--Renner relation, 
 this condition is written as follows
\be
 m_{q} \sim \frac{f_{\pi}^2/2\Si}{L^2} \label{Ec} \,\, .
\ee
 The authors of Ref.\cite{chdis,Th} suggested that Thouless energy in QCD 
 is the right hand side quantity of Eq.(\ref{Ec}).
 If relevant energy scale exceeds Thouless energy, we should expect that 
 the deviations from ChRMT appear, just like the case of Anderson model.
 This diffusion effects are discussed in Anderson model\cite{iwz}.
 We try to do the same thing in QCD. 

\subsection{Extended chiral random matrix model}
 We must introduce the parameter corresponding to
 the pion decay constant.
 To consider spatial dependence, the random matrix $W$ is replaced 
 by a set of mutually coupled random matrices as 
\be
 W \to \left( \begin{array}{cccc}
    W_1  & v1_l   &        &    0   \\
    v1_l & \ddots & \ddots &        \\
         & \ddots & \ddots &  v1_l   \\
     0   &        &  v1_l  &   W_n  \end{array}\right) \label{exW}
\ee
 where $W_i$ is an $l\times l$ random matrix and $v$ is a coupling constant
 \footnote{A similar chiral random hopping model was used in Ref.\cite{cond}
 in order to describe superconductor/normal-metal systems.}.
 Label $i$ means space-time coordinate.
 Coupling matrix $v1_l$ is chosen diagonal for simplicity.
 $1_{l}$ denotes an $l$ dimensional unit matrix .
 Eq.(\ref{exW}) is the expression for the one dimensional system.
 But we can consider any dimensions.
 The size of one block matrix, $l$, represents `internal' degrees of 
 freedom.
 In Anderson model, this size corresponds to an elastic scattering length.

 Janik et al. suggested that this size corresponds to 
 the constituent quark mass\cite{chdis}.
 In instanton liquid model, particularly in finite temperatures, 
 the instantons and anti-instantons forms molecules.
 A random matrix model motivated by this molecule formation is discussed 
 in Ref.\cite{wsw}.
 The size of these molecules can be considered $l$ in the present model.

 A probability distribution of the random matrices is taken to be Gaussian
\be
 P(H)\propto\exp\left(-l\Si^2\sum_{i=1}^{n}\mbox{Tr}W_{i}W_{i}^{\dag}\right)
\ee
 where
 $\Sigma$ is a parameter which corresponds to the chiral condensate.
 When $n = 1$, this model reduces to the original one.

 Partition function is calculated using auxiliary field as
\be
 Z_{RMT}&=&\int \prod_{i}^{n}dW_i\prod_{f=1}^{N_f}\mbox{det}\left(
   \begin{array}{cc}
       m_f     &  iW   \\
    iW^{\dag}  &  m_f
   \end{array}\right)
   \exp\left(-l\Si^2\sum_{i=1}^{n}\mbox{Tr}W_{i}W_{i}^{\dag}\right) 
 \nonumber\\
 &=& \int \prod_{i=1}^{n}dA_i\exp
 \left[-l\Si^2\sum_{i=1}^{n}\mbox{Tr}\,A_iA_i^{\dag}\right]
 \mbox{det}\left(
 \begin{array}{cc} 
  A^{\dag}+m_f & i\tilde{v} \\
  i\tilde{v} & A+m_f
 \end{array}\right)
\ee
 where
\be
 A+m_f &=& \left(\begin{array}{ccc}
  (A_1+m_f)\cdot 1_l &        &             0             \\ 
                            & \ddots &                           \\
              0             &        & (A_n+m_f)\cdot 1_l
 \end{array}\right)\\
 \tilde{v} &=& v \left(\begin{array}{cccc}
  0         & 1_{N_f l} &           & 0         \\ 
  1_{N_f l} & \ddots    & \ddots    &           \\
            & \ddots    & \ddots    & 1_{N_f l} \\
  0         &           & 1_{N_f l} & 0 
 \end{array}\right)
\ee
 and $A_i$ is an $N_f \times N_f$ matrix.
 This integral is evaluated with use of the saddle-point approximation.
 The saddle-point manifold for $v=0$ is $A_i=U_i/\Sigma$ where 
 $U_i$ is an $N_f \times N_f$ unitary matrix.
 We obtain the partition function as follows
\be
 Z_{RMT}&=&\int \prod_{i=1}^{n}dU_i\exp
 \left[\,\Si\,l\sum_{i=1}^{n}\mbox{Tr}\,m_q (U_i+U_i^{\dag})\right.
 \nonumber\\
 & & \qquad\qquad\qquad \left. +\Si^2 v^2 l\sum_{i=1}^{n-1}
 \mbox{Tr}\,(U_{i+1}^{\dag}U_i+U_{i}^{\dag}U_{i+1})+\cdots \right] .
\label{zrmt}
\ee
 The coupling parameter $v$ is assumed to be a small quantity and 
 expanded up to the second power.
 This must correspond to the partition function of the nonlinear sigma model,
\be
 Z_{NLS}&=&\int \prod_{i} dU_{i}\exp\left[\,\frac{\Si}{2}\,l\sum_{i=1}
 \mbox{Tr}\,m_q (U_i+U_i^{\dag})\right. \nonumber\\
 & & \qquad\qquad\qquad\left. +\frac{f_{\pi}^{2}}{4l}\sum_{i=1}
 \mbox{Tr}\,(U_{i+1}^{\dag}U_i+U_{i}^{\dag}U_{i+1})+\cdots \right] .
\label{znls}
\ee
 Here, we set the lattice spacing $a=1$.
 Comparing Eq.(\ref{zrmt}) with Eq.(\ref{znls}), we find 
 the coupling parameter $v$ is related with the pion decay constant as
\be
 v^2=\frac{f_{\pi}^2}{2\Si^2 l^2} .
\ee
 In the original ChRMT, the pion decay constant is treated as an 
 infinite quantity.
 We note that $v$ is small but $f_{\pi}$ is a large number.
 This means $l$ is a large number.
 In the original model, the matrix size must be taken infinity 
 in order to get the universal results.
 In the present case, the size of one block matrix $l$ must be taken infinity.
 On the other hand, it is not necessary to take an infinite limit 
 of the number of the block $n$.

 \section{Supersymmetry method}
 To calculate the level density, we use supersymmetry method\cite{ef,vwz}.
 We use the notations and conventions in Ref.\cite{vwz}.
 By this method, calculation of ensemble average can be performed easily.
 However at present this method cannot be applied for unquenched calculation
 due to the presence of the fermion determinant. 
 We use quenched approximation for this reason. 
 In original ChRMT, calculation using this method is performed by
 the authors of Ref.\cite{ast,bhz,jsv1,gw,jsv2}.
 
 Level density can be expressed as 
\be
 \rho(E)=-\frac{1}{\pi}\mbox{Im}\,\mbox{tr}\,G^{(R)}(E),
\ee
 where the (retarded) Green function is defined as
\be
 G^{(R)}(E)=\frac{1}{E^{+}-H}.
\ee
 $H$ is a $2nl\times 2nl$ matrix.
 This is derived from the generating function:
\be
 Z(E,J)&=&\frac{\mbox{det}(E^{+}-H+J)}{\mbox{det}(E^{+}-H-J)} \nonumber\\
 &=& \int d\psi \exp\left[i\psi^{\dag}(E^{+}-H-Jk)\psi\right] \label{Z}\\
 & & \qquad k=\left(\begin{array}{cc} 1&0 \\ 0&-1 \end{array}\right)
\ee
 where $\psi$ is the $4nl$-component graded vector,
\be
 \psi^{T}=\left(\phi(+)^{T},\phi(-)^{T},\chi(+)^{T},\chi(-)^{T}\right) .
\ee
In the above expression, 
 $\phi(\pm)$ is the $nl$-component bosonic vector and 
 $\chi(\pm)$ is the $nl$-component fermionic vector,
\be
 \phi^{T}(\pm)&=&\left(\phi_{1}^{T}(\pm),\ldots,\phi_{n}^{T}(\pm)\right)
 \nonumber\\
 &=&\left(\phi_{11}(\pm),\ldots,\phi_{1l}(\pm),\ldots,
 \phi_{n1}(\pm),\ldots,\phi_{nl}(\pm)\right)
\ee
\be
 \chi^{T}(\pm)&=&\left(\chi_{1}^{T}(\pm),\ldots,\chi_{n}^{T}(\pm)\right)
 \nonumber\\
 &=&\left(\chi_{11}(\pm),\ldots,\chi_{1l}(\pm),\ldots,
 \chi_{n1}(\pm),\ldots,\chi_{nl}(\pm)\right)
\ee
 where the symbol $\pm$ expresses the chiral structure of the matrix.

 Taking derivatives with respect to the external field $J$, 
 we can get the Green function.
 Generating function is normalized to unity when $J=0$ and 
 it is easy to calculate the ensemble average.

 In Eq.(\ref{dirac}), chiral basis is used.
 We perform a unitary rotation as a matter of convenience and get  
\be
 H&=& \left( \begin{array}{cc} \omega_1 & -i\omega_2 \\
 i\omega_2 & -\omega_1 \end{array}\right) .
\ee
 This basis is used in Ref.\cite{ast}.
 We follow their way and notations.

 In this basis, Dirac operator matrix is written as
\be
 & &H=\left(\begin{array}{cc}
  \omega_1+V & -i\omega_2 \\
  i\omega_2 & -\omega_1-V
 \end{array}\right) \\
 & &\quad \omega_1=\left(\begin{array}{ccc}
  \omega_{11} &        &    0        \\
              & \ddots &             \\
       0      &        & \omega_{1n}
 \end{array}\right),\qquad \omega_2=\left(\begin{array}{ccc}
  \omega_{21} &        &    0        \\
              & \ddots &             \\
       0      &        & \omega_{2n}
 \end{array}\right)\nonumber\\
 & &\quad V=\left(\begin{array}{cccc}
    0   &  v1_l  &        &  0   \\
   v1_l &    0   & \ddots &      \\
        & \ddots & \ddots & v1_l \\
    0   &        &  v1_l  &  0   
 \end{array}\right),\nonumber
\ee
 where $\omega_{1i},\omega_{2i}$ is an $l\times l$ matrix and
 $W_i=\omega_{1i}+i\omega_{2i}$.

 Performing the ensemble average over the random matrix,
 Eq.(\ref{Z}) leads to the following expression\cite{ast},
\be
 \left<Z(E,J)\right>&=&\int dW Z(E,J)
 \exp\left(-l\Si^2\sum_{i=1}^{n}\mbox{Tr}W_{i}W_{i}^{\dag}\right)\nonumber\\
 &=&\int d\psi \exp\left[-\frac{1}{4l\Si^2}
 \sum_{i=1}^{n}\trg\tilde{A}^{2}(i)+i\psi^{\dag}(E^{+}-V-Jk)\psi\right] .
 \label{Atilde}
\ee
 Here, $\tilde{A}(i)$ is a $4\times 4$ graded matrix,
\be
 & &\tilde{A}(i)=\frac{1}{\sqrt{2}}
 \left(A_z(i)-\Sigma_y A_z(i)\Sigma_y\right)\\
 & & \quad A_z(i)=\Sigma_z^{1/2}\sum_{\mu=1}^{l}
 \psi_{i\mu}\psi_{i\mu}^{\dag}\Sigma_z^{1/2}
\ee
where 
 $\psi_{i\mu}$ is a 4 dimensional graded vector
\be
 \psi_{i\mu}^{T}=\left(\phi_{i\mu}(+),\chi_{i\mu}(+),
 \phi_{i\mu}(-),\chi_{i\mu}(-)\right)
\ee
 and $\Sigma_z (\Sigma_y)$ is the $4\times 4$ Pauli matrix
 $\sigma_z \otimes 1_2$ ($\sigma_y \otimes 1_2$).

 Next, we carry out the Hubbard-Stratonovitch transformation.
 Eq.(\ref{Atilde}) is expressed with use of the auxiliary variable $Q$ as 
\be
 \left<Z(E,J)\right>&=&\int dQ d\psi \exp
 \left[-\frac{l}{2}\sum_{i=1}^{n}\trg Q^2(i)-
 \frac{i}{\sqrt{2}\Si}\sum_{i=1}^{n}\trg Q(i)\tilde{A}(i)\right.\nonumber\\
 & &\qquad\qquad\qquad \left.+i\psi^{\dag}(E^{+}-V-Jk)\psi\right]  .
\ee
 A $4\times 4$ graded matrix $Q$ is required to have 
 the same symmetry as $\tilde{A}$.
 Since $\tilde{A}$ anticommutes with $\Sigma_y$,
 we must impose the constraint
\be
 \left\{Q(i),\Sigma_y\right\}=0.
\ee
 As a consequence, $\trg Q(i)\tilde{A}(i)=\sqrt{2}\,\trg Q(i)A_z(i)$. 

 Changing the integration variable to $\psi\to\Sigma_{z}^{1/2}\psi$ and 
 performing the integral over $\psi$, 
 the generating function is expressed as follows
\be
 \left<Z(E,J)\right>&=&\int dQ \exp\left[-\frac{l}{2}
 \sum_{i=1}^{n}\trg Q^2(i)-l\sum_{i=1}^{n}\trg\mbox{ln}
 \left(E^{+}\Si_z-\frac{1}{\Si}Q(i)\right)\right.\nonumber\\
 & & +lv^2\sum_{i=1}^{n}\trg\mbox{ln}
 \left(E^{+}\Si_z-\frac{1}{\Si}Q(i)\right)^{-1}
 \left(E^{+}\Si_z-\frac{1}{\Si}Q(i+1)\right)^{-1}+\cdots\nonumber\\
 & & +\left.\sum_{i=1}^{n}\trg\mbox{ln}
 \left(E^{+}\Si_z-\frac{1}{\Si}Q(i)\right)^{-1}Jk+\cdots\right]\label{ZQ} .
\ee
 where $v$ is treated as a small parameter and expanded in a power series.
 The external field  $J$ is a diagonal matrix.
\section{Microscopic spectral density}
\subsection{Saddle-point}
 We apply a saddle-point approximation to Eq.(\ref{ZQ}).
 Since parameter $v$ is a small quantity, 
 we consider the saddle-point for $v=0$.
 The saddle-point equation is 
\be
 -Q(i)+\frac{1}{E^{+}\Si\Si_z-Q(i)}=0 .
\ee
 For $E\sim{\cal O}(l^0)$, the saddle-point manifold
 consists only of the single point
\be
 \qquad Q(i)=\Si_z\left(\frac{E\Si}{2}
 -i\sqrt{1-\left(\frac{E\Si}{2}\right)^2}\right),
\ee
 where minus sign in front of the square root is chosen so that 
 it is consistent with the sign of the infinitesimal imaginary part in $E^{+}$.
 Level density is expressed as follows
\be
 \left<\rho(E)\right>&=&-\,\frac{1}{2\pi}\ \mbox{Im}\,\mbox{tr}\,
 \left.\frac{\partial}{\partial J}
 \ \left<Z(E,J)\right>\right|_{J=0}\nonumber\\
 &=& -\frac{\Sigma l}{2\pi}\int dQ \sum_{i=1}^{n}\trg k\Sigma_z Q(i)
  \exp\left[-\frac{l}{2}\sum_{i=1}^{n}\trg Q^2(i)+\cdots\right] .
\ee
 Using the saddle-point approximation,
 we can get the familiar semicircle law
\be
 \left<\rho(E)\right>=
 \frac{1}{\D}\sqrt{1-\left(\frac{\pi E}{4nl\D}\right)^2}
\ee
 where $\Delta=\pi/2nl\Sigma$.

 For $E\sim{\cal O}(l^{-1})$, saddle-point equation becomes
\be
 Q^2(i)=-1.
\ee
 This is not a single point and 
 we must perform the integration over the saddle-point manifold. 

 Level density is expressed as follows
\be
 & &\left<\rho(E)\right>=-\frac{\Si l}{2\pi}\mbox{Im}
 \int_{Q^2=-1}dQ \sum_{i=1}^{n}\trg k\Si_z Q(i)\nonumber\\
 & &\qquad\qquad\times\exp\left[-lE\Si\sum_{i=1}^{n}\trg Q(i)\Si_z
 +l\Si^2 v^2\sum_{i=1}^{n}\trg Q(i)Q(i+1)+\cdots\right] .
\label{rhoe} 
\ee
 To proceed further, we must parametrize the saddle-point manifold.
 In Ref.\cite{ast}, the $4\times 4$ graded matrix $Q$ is parametrized as
\be
 Q &=&T_{0}^{-1}(-i\Si_z)T_{0}\\
 T_{0}&=&\left(\begin{array}{cc}
 (1+t^2)^{1/2} & it \\
 -it & (1+t^2)^{1/2}
 \end{array}\right),
\ee
 where $2\times 2$ graded matrix $t$ is given by
\be
 t=\left(\begin{array}{cc}
 a & \rho^{*} \\
 \rho & i b
 \end{array}\right).\label{t}
\ee
 We omit the indices $i$ in $Q(i)$ for the sake of simplicity of expressions. 
 Here, $a$ and $b$ are real bosonic variables and
 $\rho$ and $\rho^{*}$ are fermionic variables.
 Ref.\cite{ast} uses a diagonal parametrization of $t$, 
\be
 & & t=u t_0 u^{-1} \label{tdiag}\\
 & & \quad t_0=\left(\begin{array}{cc}
  \mu_1 & 0 \\ 0 & i \mu \end{array}\right),
 \quad u=\exp\left(\begin{array}{cc}
  0 & \xi \\ \xi^* & 0 \end{array}\right).
\ee

 In this parametrization, measure $dQ$ is calculated as
\be
 dQ &=& d\theta_1 d\theta \frac{d\xi d\xi^{*}}
 {2\pi}\frac{1}{2(\cosh\theta_1 \cos\theta
 -i\sinh\theta_1\sin\theta-1)} \\
 &=& da db \frac{d\rho d\rho^{*}}
 {2\pi}\left[1-\frac{1}{2}(\trg\,t)^2+\cdots\right]
\label{dQii}
\ee
 where 
\be
 \mu_1 &=& \sinh\frac{\theta_1}{2}\\
 \mu   &=& \sin\frac{\theta}{2}.
\ee

 In diagonal parametrization, the measure $dQ$ is expressed compactly.
 In the exact calculation for the original ChRMT, this parametrization is used.
 But we cannot do exact calculation in the present case 
 because we must treat all the nonzero modes.
 Calculation becomes very difficult and we must rely on perturbation.
 The expression of $dQ$ in the form of Eq.(\ref{dQii}) is suitable for 
 perturbative calculation.
 \subsection{Perturbative calculation}
 We take the continuum limit of Eq.(\ref{rhoe}).
 The continuum limit for a $d$ dimensional system is done by 
 the following replacement
\be
 n\to n^d,\quad l\to l^d,\quad \sum_{i=1}^{n}\to
 \frac{1}{l^d}\int d^dx,\quad Q(i+1)-Q(i)\to l\nabla Q(x) .
\ee
 The level density is expressed as
\be
 \left<\rho(E)\right>&=&-\frac{\Si}{2\pi}\mbox{Im}\int [dQ]
 \left(\int d^d x\trg\left(k\Si_z Q(x)\right)\right)\nonumber\\
 & &\times\exp\left[-\int d^d x\left(\Si E\trg Q(x)\Si_z
 -\frac{f_{\pi}^2}{4}\trg Q(x)\nabla^2Q(x)\right)\right].\label{rhoQ}
\ee
 Here, we used the relation $v^2=f_{\pi}^2/2\Sigma^2 l^2$.

 Eq.(\ref{rhoQ}) is expressed by $2\times 2$ graded matrices $t(x)$ as 
\be
 \left<\rho(E)\right>&=&\frac{2\Si}{\pi}\mbox{Re}\int [dt]
 \left\{\int d^d x(1+\trg kt^2(x))\right\}\nonumber\\
 & & \times\exp\left[-\int d^d x\left(-4i\Si E \trg t^2(x)
 -2f_{\pi}^2\trg t(x)\nabla^2 t(x)+\cdots\right)\right]\nonumber\\
 & & \times\left\{1-\frac{1}{2l^d}\int d^d x(\trg t(x))^2
 +\frac{1}{8l^{2d}}\int d^d x d^d y
 (\trg t(x))^2(\trg t(y))^2\right.\nonumber\\
 & &\qquad \left.+\frac{1}{4l^{d}}\int d^d x\trg t(x)\trg t^3(x)
 +\cdots\right\} .
\ee
 We perform the perturbative calculation.
 Keeping the bilinear terms in the exponential, 
 the other terms are expanded in a power series of $t$ and we get 
\be
 & &\left<\rho(E)\right>=\frac{2\Si(nl)^d}{\pi}\mbox{Re}
 \left<\left<1+\frac{1}{(nl)^d}\int d^d x\trg kt^2(x)
 -\frac{1}{2l^d}\int d^d x(\trg t(x))^2\right.\right.\nonumber\\
 & &-\frac{1}{2(nl)^dl^d}\int d^dxd^dy\trg kt^2(x)(\trg t(y))^2
 +\frac{1}{8l^{2d}}\int d^dxd^dy(\trg t(x))^2(\trg t(y))^2\nonumber\\
 & &\left.\left.+\frac{1}{4l^{d}}\int d^d x\trg t(x)\trg t^3(x)
 +2f_{\pi}^2\int d^dx \trg(t(x)\nabla t(x))^2 +\cdots\right>\right>\\
 & &\left<\left<\cdots\right>\right>=\int [dt](\cdots)
 \exp\left[-\int d^d x\left(-2f_{\pi}^2\trg t(x)\nabla^2 t(x)
 -4i\Si E \trg t^2(x)\right)\right] .
\ee
 The calculation is just a Gaussian integral and can be performed easily
 with help of the following identity for the arbitrary graded matrix $A$,$B$:
\be
 & &\left<\left<\trg At(x)Bt(x)\right>\right>=\Pi(E;x,y)\trg A\trg B\\
 & &\left<\left<\trg At(x) \trg Bt(x)\right>\right>=\Pi(E;x,y)\trg AB,
\ee
 where $\Pi(E;x,y)$ is the fundamental propagator of this model,
\be
 & &\Pi(E;x,y)=\sum_k \Pi(E,k)\mbox{e}^{ik \dot (x-y)}\\
 & &\Pi(E,k)=\frac{1}{4\pi(\frac{D}{\D}k^2-i\frac{E}{\D})}\\
 & &k_i=\frac{\pi n_i}{nl}\ (n_i=0,\pm 1,\ldots),\quad
 \D=\frac{\pi}{2\Si(nl)^d},\quad D=\frac{f_{\pi}^2}{2\Si} .
\nonumber
\ee
 The perturbative calculation is carried out and 
 we obtain the following expression
\be
 \left<\rho(E)\right>=\frac{1}{\D}\left\{1-\frac{n^d}{8\pi^2}
 \sum_{k}\mbox{Re}\,\frac{1}{(\frac{D}{\D}k^2-i\frac{E}{\D})^2}
 +\cdots\right\} .
\label{rho}
\ee
 In this calculation, we treat the propagator as a small quantity because
 the pion decay constant is a large quantity.
 This is correct for the nonzero modes but not correct for the zero mode 
 because the term $Dk^2/\D$ vanishes at $k=0$.
 As a result, oscillations due to level repulsion are not found.
 This is because we use the perturbation for all modes.
 Hence, for small energy regions, the above expression is no longer 
 valid due to the divergence of $k=0$ mode at $E = 0$.
 In these regions there are general arguments that 
 the microscopic spectral density should take the 
 universal form\cite{univ}.

 \section{Two-point level correlation function}
 Two-point level correlation function is calculated in the same way.
 This function can be derived from the ensemble average of 
 the product of the generating functions as 
\be
 W(E_1,E_2)&=&\left<\mbox{tr} G^{(R)}(E_1)\mbox{tr} G^{(R)}(E_2)\right>
 \nonumber\\
 &=& \frac{1}{4}\mbox{tr}\left(\frac{\partial}
 {\partial J^{(1)}}\right)\mbox{tr}\left(\frac{\partial}
 {\partial J^{(2)}}\right)\left<Z(E_1,J^{(1)})Z(E_2,J^{(2)})\right> .
\ee
 In ChRMT, we have the following identity 
\be
 \mbox{tr} G^{(A)}(E)=-\mbox{tr} G^{(R)}(-E) .
\ee
 Here, $G^{(A)}(E)$ is the advanced Green function.
 With use of this identity, the two-point level correlation function is 
 expressed as follows
\be
 \left<\rho(E_1)\rho(E_2)\right>&=&-\frac{1}{(2\pi)^2}
 \left[W(E_1,E_2)+W(-E_1,E_2)\right. \nonumber\\
 & & \qquad\qquad +\left. W(E_1,-E_2)+W(-E_1,-E_2)\right] .
\ee

 Now let us turn to the calculation of $W(E_1,E_2)$.
 The generating function is expressed with use of graded matrices 
 and vectors as 
\be
 Z(E_1,J^{(1)})Z(E_2,J^{(2)})=\int d\hat{\psi}\exp
 \left[i\hat{\psi}^{\dag}(\hat{E}-H-\hat{J}k)\hat{\psi}\right]
\ee
 where 
\be
 \hat{\psi}=\left(\begin{array}{c}\psi^{(1)}\\
 \psi^{(2)}\end{array}\right),\quad \hat{E}=
 \left(\begin{array}{cc} E_1 & 0 \\ 0 & E_2 \end{array}\right),
 \quad \hat{J}=\left(\begin{array}{cc}J^{(1)} & 0 \\
  0 & J^{(2)} \end{array}\right) .
\ee
 As in the previous section, the ensemble average and 
 Hubbard-Stratonovitch transformation is performed and we obtain 
\be
 & &\left<Z(E_1,J^{(1)})Z(E_2,J^{(2)})\right>
 =\int dQ\exp\left[-\frac{l}{2}\sum_{i=1}^{n}\trg Q^{2}(i)\right. \nonumber\\
 & & -l\sum_{i=1}^{n}\trg \ln (\hat{E}\Si_z
 -\frac{1}{\Si}Q(i))
 +lv^2\sum_{i=1}^{n}\trg (\hat{E}\Si_z
 -\frac{1}{\Si}Q(i))^{-1}(\hat{E}\Si_z
 -\frac{1}{\Si}Q(i+1))^{-1}\nonumber\\
 & &\left.+\sum_{i=1}^{n}\trg
 (\hat{E}\Si_z-\frac{1}{\Si}Q(i))^{-1}\hat{J}_{ii}'k+\cdots \right].
\ee
 Here, $Q$ is an $8\times 8$ graded matrix.
 For $E\sim {\cal O}(l^{-1})$, the saddle point is $Q(i)^{2} =-1$.
 Parametrization is as follows\cite{ast}
\be
 Q&=&T_{0}^{-1}(-i\Si_z)T_0\\
  & & T_0=T_{u}T_{ch} .
\nonumber
\ee
 An $8\times 8$ graded matrix $T_0$ is separated by the 
 chiral rotation part $T_{ch}$ and the unitary rotation part $T_{u}$ as
\be
 T_{ch}&=&\left(\begin{array}{cccc}
  (1+t_{1}^{2})^{1/2} & 0 & it_1 & 0 \\
  0 & (1+t_{2}^{2})^{1/2} & 0 & it_2 \\
  -it_1 & 0 & (1+t_{1}^{2})^{1/2} & 0 \\
  0 & -it_2 & 0 & (1+t_{2}^{2})^{1/2}
 \end{array}\right)\\
 T_{u}&=&\left(\begin{array}{cccc}
  (1+t_{12}t_{21})^{1/2} & 0 & 0 & it_{12} \\
  0 & (1+t_{21}t_{12})^{1/2} & it_{21} & 0 \\
  0 & it_{12} & (1+t_{12}t_{21})^{1/2} & 0 \\
  -it_{21} & 0 & 0 & (1+t_{21}t_{12})^{1/2}
 \end{array}\right).
\ee
 Here, $2\times 2$ graded matrix $t$ is defined as 
\be
 & & t_1=\left(\begin{array}{cc}a_1 & \rho_{1}^{*} \\
 \rho_1 & ib_1 \end{array}\right),\quad t_2=
 \left(\begin{array}{cc}a_2 & \rho_{2}^{*} \\
 \rho_2 & ib_2 \end{array}\right) \\
 & & t_{12}=\left(\begin{array}{cc} c & i\eta \\
 \sigma & id \end{array}\right),\quad t_{21}=
 \left(\begin{array}{cc} c^{*} & \sigma^{*} \\
 -i\eta^{*} & id^{*}\end{array}\right).
\ee
 The chiral rotation is due to the chiral structure of the matrix
 and the unitary rotation is due to the calculation
 of the two-point level density.
 The former is the same as the calculation of the one-point level density in 
 section 4
 and the latter is the same as the calculation of the two-point level density 
 in the standard (non-chiral) RMT.

 Taking derivative with respect to $J$, we can obtain $W(E_1,E_2)$ as
\be
 W(E_1,E_2)&=&\frac{l^2 \Si^2}{4}\int_{Q^2=-1} dQ 
 \sum_{i,j}\trg \left(k\Si_z\frac{1+\Lambda}{2}Q(i)\right)
 \trg \left(k\Si_z\frac{1-\Lambda}{2}Q(i)\right)\nonumber\\
 & &\times\exp\left[-l\Si\sum_{i=1}^{n}\trg \hat{E}^{+}\Si_z Q(i)
 +lv^2\Si^2 \sum_{i=1}^{n}\trg Q(i)Q(i+1)\right] \label{w12}
\ee
 where
\be
 \Lambda =\mbox{diag}\left(1,-1,1,-1\right).
\ee
 In this representation, $8\times 8$ graded matrix is aligned 
 as the following block structure: chiral, two-point, boson-fermion,
\be
 M&=&\left(\begin{array}{cc}
  M(++) & M(+-) \\ 
  M(-+) & M(--) \end{array}\right)\\
 M(++)&=&\left(\begin{array}{cc}
  M^{(11)}(++) & M^{(12)}(++) \\
  M^{(21)}(++) & M^{(22)}(++) \end{array}\right)\\
 M^{(11)}(++)&=&\left(\begin{array}{cc}
  M_{BB}^{(11)}(++) & M_{BF}^{(11)}(++) \\
  M_{FB}^{(11)}(++) & M_{FF}^{(11)}(++) \end{array}\right).
\ee
 We calculate the integral of Eq.(\ref{w12}) perturbatively.
 The measure $dQ$ is separated into the chiral part and
 unitary (two-point) part: $dQ=d\mu(T_{ch})d\mu(T_u)$.
 The chiral part is the same as the previous section.
 Jacobian of the unitary part is equal to unity if 
 we take the matrix elements of $t_{12},t_{21}$
 as the integral variables.
 Then we get 
\be
 dQ=dt_1dt_2\left(1-\frac{1}{2}\trg t_1^2-\frac{1}{2}\trg t_2^2
 +\cdots\right)dt_{12}dt_{21} .
\ee
 The calculation is carried out in the same way as the previous section
 and yields 
\be
 W(E_1,E_2)&=&-4(nl)^{2d}\Si^2 +8\Si^{2}n^{d}(nl)^{2d}
 \sum_{k}\left(\Pi^2(E_1,k)+\Pi^2(E_2,k)\right)\nonumber\\
 & & -16\Si^{2}(nl)^{2d}\sum_{k}\Pi^2\left(\frac{E_1+E_2}{2},k\right)+\cdots .
\ee
 Two-point level density is expressed as 
\be
 & &\left<\rho(E_1)\rho(E_2)\right>-\left<\rho(E_1)\right>
 \left<\rho(E_2)\right> \nonumber\\
 & &=\frac{1}{\D^2}\sum_{k}\left[\Pi^2\left(E,k\right)
 +\Pi^2\left(-E,k\right)+\Pi^2\left(\omega/2,k\right)
 +\Pi^2\left(-\omega/2,k\right)\right]+\cdots, \label{2pt}
\ee
 where $E=(E_1+E_2)/2$ and $\omega=E_1-E_2$.
 $E$-dependent terms are characteristic in ChRMT.
 Comparing this expression with the usual (non-chiral) RMT,
 we confirm that Thouless energy in QCD is 
\be
 E_c=\frac{f_{\pi}^2/2\Sigma}{L^2}.
\ee

 We can calculate number variance from the two-point level density.
 This is defined as
\be
 \Sigma^2(z)=\int_{-z/2}^{z/2} dx dy \D^2 
 \left(\left<\rho(E_1)\rho(E_2)\right>
 -\left<\rho(E_1)\right>\left<\rho(E_2)\right>\right),
\label{nv}
\ee
 where $x=E_1/\D$ and $y=E_2/\D$.
 This is the measure of fluctuations of the level number
 in the finite level interval $z$.

 Integration is performed and we obtain
\be
 \Sigma^2(z)= \frac{1}{\pi^2}\sum_k \mbox{ln}
 \left(1+\frac{z^2}{4(\frac{D}{\D}k^2)^2}\right).
\ee
 We note that the $E$-dependent terms and $\omega$-dependent terms give 
 the same contributions to the number variance.
 Eq.(\ref{2pt}) does not treat the zero mode exactly.
 We expect, however, that this approximation does not 
 change the behavior of $\Sigma^{2}(z)$ drastically.
 This is because it is known from the results of the usual RMT
 that what is missing in the perturbative result of the two-point 
 level density is the oscillating behavior of the exact result
 and the integration in Eq.(\ref{nv}) cancels these oscillations.
 Taking the continuum limit, we obtain
\be
 \Sigma^2(z)&=& \frac{1}{\pi^2}\left(\frac{\D z}{2D}\right)^{d/2}
 \frac{V}{\pi^d}\int d^dk' \ \mbox{ln}
 \left(1+\frac{1}{k'^4}\right) \\ 
 &\propto& z^{d/2}. \nonumber 
\ee
 We find that the number variance depends on
 the dimensionality of the system.
 This is very different from the original ChRMT 
 where $\Sigma^2(z)\sim\mbox{log}z$ for large $z$.
 The similar behavior is find in Ref.\cite{as} for usual (non-chiral) RMT
 and interpreted as change from level correlation of RMT 
 to those of non-correlated energy levels
 (Poisson statistics : $\Sigma^2(z)\sim z$).
 This is consistent with intuitive discussion.

 \section{Two-point correlation function}
 The spatial dependence, which we introduced into the original ChRMT 
 in this paper, enables us to 
 calculate the correlation between two spatial point.
 By formal replacement of the energy $E^+$ to $im$, the Green function 
 $G^{(R)}(E)$ becomes a quark propagator.
 The ensemble average of a matrix element 
\be
 G^{(R)}(im;x,y,\mu,\nu)=\left<x,\mu\left|\frac{1}{im-H}\right|y,\nu\right>
\ee
 can be calculated from the generating function as
\be
 \left<G(im;x,y,\mu,\nu)\right>=
 \frac{1}{2}\frac{\partial}{\partial J_{xy,\mu\nu}}\left<Z(im,J)\right>.
\ee
 The external field $J$ is not diagonal in this case.
 Label $x$, $y$ represents the position of block matrices.
 Label $\mu$,$\nu$ represents the `internal' degrees of freedom 
 which are described and by random matrices.
 We get $\left<G(im;x,y,\mu,\nu)\right>=0$ if $\vert x-y\vert > l$.
 This behavior means the size $l$ is a reciprocal of 
 the constituent quark mass.
 This is consistent with the discussion of the Ref.\cite{chdis}.

 We can also consider `meson' propagator in this model.
 This is the product of the Green functions and can be calculated from
 the generating function as
\be
 & &\left<\mbox{tr}\sum_{\mu,\nu}G(im;x,y,\mu,\nu)G(im;y,x,\nu,\mu)\right>= 
 \nonumber\\
 & &\frac{1}{4}\sum_{\mu\nu}\left(
 \frac{\partial}{\partial J_{xy\mu\nu}(+)}
 \frac{\partial}{\partial J_{yx\nu\mu}(+)}+
 \frac{\partial}{\partial J_{xy\mu\nu}(+-)}
 \frac{\partial}{\partial J_{yx\mu\nu}(-+)} \right.\nonumber\\
 & &\qquad\left. +\frac{\partial}{\partial J_{xy\mu\nu}(-+)}
 \frac{\partial}{\partial J_{yx\mu\nu}(+-)}+
 \frac{\partial}{\partial J_{xy\mu\nu}(-)}
 \frac{\partial}{\partial J_{yx\mu\nu}(-)} \right)\left<Z(E,J)\right>
 \label{sigma}
\ee
 and
\be
 & &\left<\mbox{tr}\sum_{\mu,\nu}G(im;x,y,\mu,\nu)\gamma_5
 G(im;y,x,\nu,\mu)\gamma_5\right>= \nonumber\\
 & &\frac{1}{4}\sum_{\mu\nu}\left(
 \frac{\partial}{\partial J_{xy\mu\nu}(+)}
 \frac{\partial}{\partial J_{yx\nu\mu}(-)}+
 \frac{\partial}{\partial J_{xy\mu\nu}(+-)}
 \frac{\partial}{\partial J_{yx\mu\nu}(+-)} \right.\nonumber\\
 & &\qquad\left. +\frac{\partial}{\partial J_{xy\mu\nu}(-+)}
 \frac{\partial}{\partial J_{yx\mu\nu}(-+)}+
 \frac{\partial}{\partial J_{xy\mu\nu}(-)}
 \frac{\partial}{\partial J_{yx\mu\nu}(+)} \right)\left<Z(E,J)\right> .
 \label{pi}
\ee
 Here, trace is taken over the chiral degrees of freedom.
 Summation over $\mu$,$\nu$ is taken
 because this degrees of freedom is described by random matrices.
 Eq.(\ref{sigma}) corresponds to scalar (sigma) propagator and 
 Eq.(\ref{pi}) corresponds to pseudo scalar (pion) propagator
 \footnote{We must insert the isospin matrix
 but this is irrelevant in this calculation.}.
 We consider the long range correlation between the point $x\ne y$.
 In the present basis, matrix $\gamma_5$ is expressed as
\be
 \gamma_5=\left(\begin{array}{cc}0 & -1 \\ -1 & 0\end{array}\right).
\ee
 Eq.(\ref{sigma}) and (\ref{pi}) are calculated perturbatively and we obtain 
\be
 & &\left<\mbox{tr}\sum_{\mu,\nu}G(im;x,y,\mu,\nu)G(im;y,x,\nu,\mu)\right>
 \nonumber\\
 & &\qquad =64\Si^2 l^d \int d^dz\Pi(im;x,y)\Pi(im;y,z)\Pi(im;z,x)+\cdots
\ee
\be
 & &\left<\mbox{tr}\sum_{\mu,\nu}G(im;x,y,\mu,\nu)\gamma_{5}
 G(im;y,x,\nu,\mu)\gamma_5\right>=-16l^{2d}\Si^2 \Pi(im;x,y)+\cdots
 \label{pimode}
\ee
 where
\be
 \Pi(im;x,y)&=&\sum_{k} \Pi(im,k)\exp\left(ik(x-y)\right)\\
 \Pi(im,k)&=&\frac{1}{4\pi\left(\frac{D}{\D}k^2+\frac{m}{\D}\right)}
 \nonumber\\
 &=&\frac{1}{4\pi\frac{D}{\D}(k^2+m_{\pi}^2)}.
\ee
 Here $\Pi(im;k)$ is obtained from $\Pi(E;k)$ 
 by the replacement of $E$ by $im$.
 The above equation shows that $\Pi(im;x,y)$ corresponds to 
 the pion propagator.
 We find that the scalar propagator can be neglected
 by comparing with the pseudo scalar propagator.
 Pseudo-scalar mode is important.
 This is consistent with the NG theorem.
 As was shown in Section 2.2, the present model is equivalent to 
 the nonlinear sigma model.
 Hence it is reasonable that the pion propagator appears as 
 a fundamental ingredient in the calculation of various correlation 
 functions.
 The calculation of the two-point correlation function was also 
 done in Ref.\cite{pqch} with use of 
 partially quenched chiral perturbation theory.

 $\Pi(E;x,y)$ is a propagator of the fundamental 
 mode in the perturbative calculation.
 In Anderson problem, this mode is called diffuson.
 Eq.(\ref{pimode}) shows the pion mode in QCD plays the same role as
 the diffuson mode in Anderson model.
 The fundamental mode appealing 
 in the chiral perturbation theory, that is, pion, 
 corresponds to the fundamental mode 
 appealing in the Anderson model, diffuson.
\section{Conclusions}
 In this paper, we examined the level correlations of QCD Dirac operator.
 We introduced the extended chiral random matrix model
 including the spatial dependence.

 First, we showed the partition function of this random matrix model is 
 equivalent to that of a nonlinear sigma model.  
 This equivalence has been previously shown only for the 0-dimensional case 
 without spatial dependence. 
 As far as we know, this is a first explicit derivation of nonlinear sigma
 model (with derivative terms) from the kind of chiral random matrix models. 
 
 To calculate level statistics,
 we used supersymmetry method developed by Efetov.
 Using the perturbative expansion, we got the expression of the level 
 density and two-point level correlation function.

 Comparing the expression of this function with the usual RMT,
 we confirm that the Thouless energy in QCD is $E_c=f_\pi^2/2\Si L^2$.
 Number variance was calculated from the two-point level correlation function
 and found to depend on the dimensionality of the system.
 This is very different from the standard universal result and 
 interpreted as change from level correlation of RMT 
 to those of non-correlated energy levels.

 For the first time, the correlation functions of 
 pseudo-scalar type as well as scalar type 
 between two spatial points could be calculated for the chiral 
 random matrix model.
 The previous work with use of partially quenched chiral perturbation 
 theory\cite{pqch} only gives the correlation function of pseudo-scalar type.  
 The reason why scalr type correlation function can be 
 calculated in this work is that quark propagators can be 
 defined for the chiral random matirx model.

 These functions demonstrate that 
 the pion propagator is nothing but the diffuson propagator which is
 a fundamental mode responsible for diffusion effects in Anderson model.
 Namely, pion mode plays a role of diffuson when we consider QCD vacuum 
 as a disordered medium.
 On the other hand, sigma-like mode does not appear as a fundamental 
 mode in this model.
 
\section*{Acknowledgements}
 We would like to thank T.Kunihiro for useful discussions and 
 T.Wettig and J.J.M.Verbaarschot for useful comments.
 K.T is supported by Research Fellowships of 
 the Japan Society for the Promotion of Science for Young Scientists.
 
\end{document}